\documentclass[twocolumn,english,aps,prl]{revtex4}
\usepackage[T1]{fontenc}
\usepackage[latin1]{inputenc}
\usepackage{geometry}
\usepackage{graphicx}
\usepackage{amssymb}

\makeatletter

\providecommand{\LyX}{L\kern-.1667em\lower.25em\hbox{Y}\kern-.125emX\@}
\usepackage{graphicx}

\def\Tr{\mathop{\rm Tr}}
\usepackage{txfonts}

\usepackage{babel}
\makeatother
\begin{document}

\title{Communication near the channel capacity with an absence of compression:
\\
 Statistical Mechanical Approach}

\author{Ido Kanter and Hanan Rosemarin}

\affiliation{Minerva Center and the Department of Physics, Bar-Ilan University,
\\
 Ramat-Gan 52900, Israel}

\begin{abstract}
The generalization of Shannon's theory to include messages with given
autocorrelations is presented. The analytical calculation of the channel
capacity is based on the transfer matrix method of the effective 1D
Hamiltonian. This bridge between statistical physics and information
theory leads to efficient Low-Density Parity-Check Codes over Galois
fields that nearly saturate the channel capacity. The novel idea of
the decoder is the dynamical updating of the prior block probabilities
which are derived from the transfer matrix solution and from the posterior
probabilities of the neighboring blocks. Application and possible
extensions are discussed, specifically the possibility of achieving
the channel capacity without compression of the data. 
\end{abstract}


\maketitle
\begin{figure}
\includegraphics[  width=0.85\columnwidth,
  keepaspectratio]{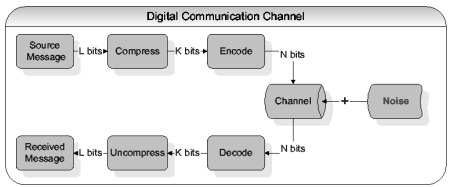}

\caption{\label{fig-dig-chnl}Digital Communication Channel}
\end{figure}

Digital communication, which is the driving force of the modern information
revolution, deals with the task of achieving reliable communication
over a noisy channel. A typical communication channel is depicted
in figure \ref{fig-dig-chnl} (where $K\leq L$ and $K\leq N$).%

Shannon, in his seminal work\cite{Shannon-48}, proved that in order
to overcome noise, redundancy must be added to the transmitted message.
The channel capacity, which is the maximal rate ($R\equiv \frac{K}{N}$)
is a function of the channel bit error rate ($f$), bit error rate
($P_{b}$) and the a priori bit probability ($P$) \begin{equation}
R=\frac{1-H_{2}\left(f\right)}{H_{2}\left(P\right)-H_{2}\left(P_{b}\right)}\label{Shannon-Capacity-eq}\end{equation}
 where $H_{2}\left(x\right)\equiv -x\log _{2}\left(x\right)-\left(1-x\right)\log _{2}\left(1-x\right)$.
The entropy, defined as the information content (in bits per symbol)
of the message, is unity for unbiased ($P=0.5$) messages and decreases
with bias ($H_{2}\left(0\right)=H_{2}\left(1\right)=0$). 

To maximize channel throughput, the message is first compressed using
algorithms (e.g., \cite{LZ78}) which approach entropy for large blocks,
and then encoded assuming a unbiased message. Typical messages exhibit
low autocorrelation coefficients (decreasing with the size of the
message)\cite{lacs}. For instance, the discrete periodic 2-point
autocorrelation coefficient of a sequence $\left\{ X_{i}\right\} $
is defined by \begin{equation}
C_{k}=\frac{1}{L}\sum _{i=1}^{L}X_{i}X_{\left(i+k\right)\: \mathbf{mod}\: L}\label{autocorrelation-coeff-eq}\end{equation}

Compressible data exhibits enhanced autocorrelation coefficients,
whose distribution has been extensively studied in numerous fields
(e.g., linguistics, DNA research, heartbeat intervals, etc.)\cite{zipf,Havlin}. 

In this paper we raise two questions: 

\begin{itemize}
\item what is the channel capacity of correlated sequences? 
\item are these bounds achievable by an encoder alone, without the preceding
compression phase? 
\end{itemize}
We address these questions by using models from statistical mechanics,
which lead to a new class of decoding algorithms. The decoder nearly
approaches the capacity expected for perfect compression, without
directly applying any compression. 

The motivation for direct transmission of the uncompressed data is
to overcome many difficulties, including: 

\begin{itemize}
\item a single bit error in a compressed block would render the entire block
useless. This is particularly important for applications where minimal
distortion can be tolerated (e.g., audio/video transmission) 
\item compression of different packets of the same length would generate
different length output messages, calling for greater complexity of
the encoder 
\item the compression/decompression add delay to each transmission, and
increase the complexity of the transmitter/receiver. 
\end{itemize}
We start by calculating the entropy of correlated bit sequences. This
is done by binning the source into $K_{o}$ bits, where $K_{o}$ is
the highest autocorrelation coefficient taken, and using the transfer-matrix
method. We proceed by integrating the results of the transfer matrix
into a new decoding algorithm, based on a Low-Density Parity-Check
Codes (LDPCC) over finite fields ($GF\left(q\right)$)\cite{LDPC-GF(q)}. 

For the sake of simplicity, we begin by demonstrating the entropy
calculation for two autocorrelation coefficients, namely, $C_{1}$
and $C_{2}$. The entropy of a set of binary sequences of length $L$
is defined by the log of the number of possible sequences ($\Omega $)
divided by $L$. Calculating $\Omega $ with the correlation constraints
($C_{1}$, $C_{2}$)\cite{sourlas}, is done by taking the trace over
all possible states of the sequence obeying the constraints imposed
as delta functions\begin{equation}
\Omega =\Tr \delta \left(\sum _{j}x_{j}x_{j+1}-C_{1}L\right)\delta \left(\sum _{j}x_{j}x_{j+2}-C_{2}L\right)\label{omega-c1-c2}\end{equation}
 where periodic boundary conditions are assumed for the indices. Using
the Fourier representation of the delta functions and rearranging
terms, one gets\begin{eqnarray}
\Omega  & = & \frac{1}{\left(2\pi \right)^{2}}\int \int _{-\infty }^{\infty }\mathbf{d}y_{1}\mathbf{d}y_{2}e^{-iL\left(y_{1}C_{1}+y_{2}C_{2}\right)}\nonumber \\
 &  & \times \Tr \prod _{j=1}^{\frac{L}{2}}e^{iB\left(y_{1},y_{2},x_{2j}\ldots x_{2j+3}\right)}\label{Omega-c1-c2-fourier}
\end{eqnarray}
 where $B$ is given by \begin{eqnarray}
B\left(y_{1},y_{2},x_{0},x_{1},x_{2},x_{3}\right) & \equiv  & \frac{y_{1}}{2}\left(x_{0}x_{1}+2x_{1}x_{2}+x_{2}x_{3}\right)\nonumber \\
 &  & +y_{2}\left(x_{0}x_{2}+x_{1}x_{3}\right)\label{Omega-c1-c2-B}
\end{eqnarray}
 Using the transfer-matrix method\cite{baxter-TM}, we group the sequence
into blocks of 2 bits, obtaining the 4x4 matrix \begin{equation}
V\left(y_{1},y_{2}\right)=\left(\begin{array}{cccc}
 e^{2y_{1}+2y_{2}} & e^{y_{1}} & e^{-y_{1}} & e^{-2y_{2}}\\
 e^{-y_{1}} & e^{-2y_{1}+2y_{2}} & e^{-2y_{2}} & e^{y_{1}}\\
 e^{y_{1}} & e^{-2y_{2}} & e^{-2y_{1}+2y_{2}} & e^{-y_{1}}\\
 e^{-2y_{2}} & e^{-y_{1}} & e^{y_{1}} & e^{2y_{1}+2y_{2}}\end{array}
\right)\label{TM-V}\end{equation}
 representing all possible interactions between neighboring blocks.
We proceed by replacing the trace with $\lambda _{_{max}}^{^{\frac{L}{2}}}$,
where $\lambda _{max}$ is the principal eigenvalue of $V\left(y_{1},y_{2}\right)$.
Using the method of Laplace integrals we obtain for the leading order\cite{prl-referee}
of $\Omega $\begin{equation}
\Omega =e^{-L\left(y_{1}^{\star }C_{1}+y_{2}^{\star }C_{2}-\frac{1}{2}\ln \lambda _{max}\left(y_{1}^{\star },y_{2}^{\star }\right)\right)}\label{Omega-c1-c2-TM}\end{equation}
 where $\left(y_{1}^{\star },y_{2}^{\star }\right)$ are the solution
of the saddle point equations. The entropy in bits is given by \begin{equation}
H\left(C_{1},C_{2}\right)=\frac{\frac{1}{2}\ln \lambda _{max}\left(y_{1}^{\star },y_{2}^{\star }\right)-y_{1}^{\star }C_{1}-y_{2}^{\star }C_{2}}{\ln 2}\label{entropy-c1-c2}\end{equation}

\begin{figure}
\includegraphics[  height=0.85\columnwidth,
  keepaspectratio,
  angle=270,
  origin=lB]{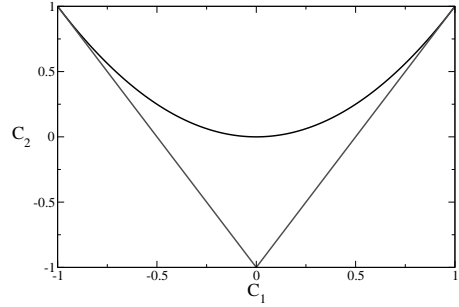}

\caption{\label{fig-c1c2-2d}Phase space for sequences with $C_{1},C_{2}$.
Outside the triangle ($\left|C_{1}\right|\geq \frac{1+\left|C_{2}\right|}{2}$)
the entropy is zero. The parabolic curve is $C_{2}=C_{1}^{2}$.}
\end{figure}

The entropy of the correlated sequences are shown in figure \ref{fig-c1c2-2d}
for all possible \{$C_{1},C_{2}$\} tuples. The area of non-zero entropy
is depicted in figure \ref{fig-c1c2-2d} between the two straight
lines, i.e., $\frac{-1-C_{2}}{2}<C_{1}<\frac{1+C_{2}}{2}$. The parabolic
line is $C_{2}=C_{1}^{2}$, where the entropy reduces to the case
of a single $C_{1}$ constraint, with the entropy $S=H_{2}\left(\frac{1+C_{1}}{2}\right)$\cite{C1-gauge}.
This is the typical case for the $C_{1}$ constraint, since $C_{2}$
can be viewed as multiplying two consecutive $C_{1}$ pairs, hence
it has the highest entropy for a given $C_{1}$. 

Note that at the boundary of the phase space the entropy falls abruptly
to zero (a first order phase transition). Another point, which might
appear counter-intuitive, is the fact that the entropy of the constraint
\{$C_{1}=A,C_{2}=B$\} is not the same as \{$C_{1}=B,C_{2}=A$\} \cite{corr-cluster}. 

The transfer-matrix solution shows that the ensemble of sequences
obeying the correlation constraints are obtained from the thermal
equilibrium solution of the 1-D Ising spin model (where the two states
of the spin correspond to the binary values of the bit $\left\{ 0,1\right\} \rightarrow \left\{ 1,-1\right\} $
\cite{sourlas}). The interaction length is the same as the correlation
length, and the interaction strength is $J_{i}=-\frac{y_{i}^{\star }}{\beta }$
(where $\beta $ is the inverse temperature). The corresponding Hamiltonian
is $H=-\frac{y_{1}^{\star }}{\beta }\sum x_{i}x_{i+1}-\frac{y_{2}^{\star }}{\beta }\sum x_{i}x_{i+2}$.
This physical observation is the key leading to our novel decoding
algorithm. 

Testing the physical mapping, we choose a pair of constraints ($C_{1},C_{2}$),
solve the transfer matrix model to obtain $y_{1}^{\star },y_{2}^{\star }$,
then select any temperature (e.g., $\beta =1$) which gives the interactions
$J_{1},J_{2}$\cite{frustrated_interactions}, and perform Monte-Carlo
simulations, and indeed the system settles into an equilibrium state
obeying the initial autocorrelation constraints. 

The division into blocks of 2 bits, amplifies the fact that interactions
affect only neighboring blocks, and the probability of finding a block
in one of its four possible states \emph{depends only on the state
of its two neighbors}. Labeling the left, center and right blocks
by $l,c,r$ respectively, one gets for the joint probability of three
consecutive blocks \begin{equation}
P\left(c,l,r\right)=\frac{S_{I}\left(c\right)q_{L}^{l}S_{L}\left(l,c\right)q_{R}^{r}S_{R}\left(r,c\right)}{\Tr _{\left\{ c,l,r\right\} }S_{I}\left(c\right)q_{L}^{l}S_{L}\left(l,c\right)q_{R}^{r}S_{R}\left(r,c\right)}\label{Block-Interaction}\end{equation}
 where\begin{eqnarray}
S_{I}\left(c\right) & = & e^{y_{1}^{\star }c_{1}c_{2}}\nonumber \\
S_{L}\left(l,c\right) & = & e^{y_{1}^{\star }l_{2}c_{1}+y_{2}^{\star }\left(c_{1}l_{1}+c_{2}l_{2}\right)}\label{Block-State-Prob}\\
S_{R}\left(r,c\right) & = & e^{y_{1}^{\star }c_{2}r_{1}+y_{2}^{\star }\left(c_{1}r_{1}+c_{2}r_{2}\right)}\nonumber 
\end{eqnarray}
 The first term in Equation \ref{Block-State-Prob} denotes the inner-block
interactions, and the other terms denote the inter-block interactions.
$q_{L}^{l}$/$q_{R}^{r}$ is the posterior probability of finding
the left/right block in a specific state given by the decoding algorithm
below, and the subscripts 1,2 denote the bit number in the block. 

The physical spin model emphasizes the dramatic difference between
the case of biased bit probabilities, equivalent to an induced homogeneous
field, where each spin is updated independently, to the case of correlations
induced by inter-spin interactions. 

The entropy of the class of correlated sequences, derived by the transfer-matrix
method, can be plugged into equation \ref{Shannon-Capacity-eq}. Using
shannon's lower bound, gives the channel capacity of sequences with
two autocorrelation coefficients \begin{equation}
R^{\star }=\frac{1-H_{2}\left(f\right)}{H_{2}\left(C_{1},C_{2}\right)-H_{2}\left(P_{b}\right)}\label{generalizd-Capacity-eq}\end{equation}

The previous formulation is easily extended to higher autocorrelation
coefficients. The procedure involves adding more delta functions to
equation \ref{omega-c1-c2}, resulting in a larger transfer matrix.
For $C_{l}$ , being the highest autocorrelation coefficient, one
gets $l$ variables ($y_{1}\ldots y_{l}$), and a transfer matrix's
dimension is $2^{l}\times 2^{l}$, which is solved numerically. The
simulation results, shown in this paper, were conducted on sequences
with 2 ($C_{1},C_{2}$), and 3 ($C_{1},C_{2},C_{3}$) autocorrelation
constraints. 

Equation \ref{Block-Interaction} is the crux of our decoding algorithm,
it enables the decoder to set prior probabilities for each block based
on the current state of its two neighbors. 

Decoding of the autocorrelated sequences is based on LDPCC which have
been shown to asymptotically nearly saturate Shannon's bound\cite{forney,richardson,KS-LDPC}.
These codes easily lend themselves to decoding blocks of $t$ bits
by moving from boolean algebra to Galois-Field ($GF\left(q\right)$,
with $q=2^{t}$). This method was originally studied\cite{LDPC-GF(q)},
as a means of increasing the degree of the nodes (connectivity of
the graph) without introducing loops, which are known to degrade the
decoder's performance. 

The decoding consists of iterating two rounds of message passing (also
known as belief-propagation), which update two probability matrices
$R_{mn}^{a},Q_{mn}^{a}$ ($a\in 0..\left(q-1\right)$ stands for the
state of the $\left(m,n\right)$ element, where $m\in \left\{ 1\ldots \frac{N}{t}\right\} $,
and $n\in \left\{ 1\ldots \frac{N+K}{t}\right\} $ is the block number
\cite{MN-Gallager}). 

\begin{itemize}
\item Updating $R_{mn}$ is based on the values of $Q_{mn}$, and the specific
parity check. This stage is left unchanged. 
\item Updating $Q_{mn}$ is based on $R_{mn}$ and the a priori probabilities
of the $q$ states of the block\begin{equation}
Q_{mn}^{a}=\alpha _{mn}P_{n}^{a}\prod _{j\in M\left(n\right)\setminus m}R_{jn}^{a}\label{vertical-pass}\end{equation}
 where $\alpha _{mn}=\frac{1}{\sum _{a=1}^{q}Q_{mn}^{a}}$ is a normalizing
constant. 
\end{itemize}
In this stage the a priori probabilities, $P_{n}^{a}$, of the block,
are obtained by marginalizing the joint probability (equation \ref{Block-Interaction})
\begin{eqnarray}
\gamma _{n}^{c} & = & S_{I}\left(c\right)\left(\sum _{l=1}^{q}q_{L}^{l}S_{L}\left(l,c\right)\right)\left(\sum _{r=1}^{q}q_{R}^{r}S_{R}\left(r,c\right)\right)\nonumber \\
P_{n}^{c} & = & \frac{\gamma _{n}^{c}}{\sum _{j=1}^{q}\gamma _{n}^{j}}\label{tm-vertical-pass}
\end{eqnarray}
 where $l/r$ denotes the state of the $n-1/n+1$ block respectively,
and $q_{L}^{l}/q_{R}^{r}$ are their posterior probabilities. The
meaning of equation \ref{tm-vertical-pass} is summing the weighted
contributions from all $q^{2}$ possible states of the two neighboring
blocks to a possible state of the block in question. Thus, in each
iteration the prior probability of a block is \emph{dynamically updated}
following its neighboring blocks. 

The decoder performs the iterations until the message is decoded (all
checks are satisfied), or one of two alternate halting criteria is
met: (a) the maximal number of iterations is reached. In this paper
the maximum was set to 500; (b) the decoder has not modified its estimate
of the message in the last 50 iterations. 

Simulations were run using the Binary Symmetric Channel (BSC), and
the construction of the check matrix ($H$) used is based on KS Codes\cite{KS-LDPC},
for $R=\frac{1}{3}$, where the non-zero elements of $H$ are randomly
drawn from the range $1\ldots \left(q-1\right)$. 

\begin{figure}
\includegraphics[  height=0.85\columnwidth,
  keepaspectratio,
  angle=270,
  origin=lB]{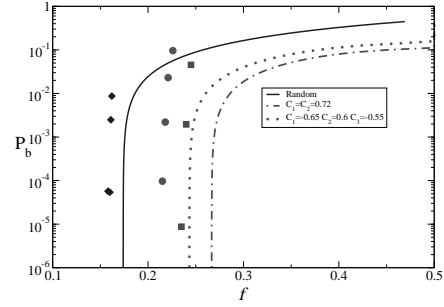}

\caption{\label{fig-pb-f-c1-c2}$P_{b}$ (decoder bit error rate) vs. $f$
(channel bit error rate) with rate $R=\frac{1}{3}$ for uncorrelated
sequences; for sequences with $C_{1}=C_{2}=0.72$; and for sequences
with $C_{1}=-0.65$, $C_{2}=0.6$, $C_{3}=-0.55$ . Results of simulations
using KS codes\cite{KS-LDPC}, with $N=10^{4}$ (9,999 for GF(8)),
averaged over at least 2,000 samples, are given for the regular algorithm
($\blacklozenge $) and for the new algorithm ($\blacksquare $ and
$\medbullet $ for the 2 and 3 autocorrelations respectively).}
\end{figure}

Figure \ref{fig-pb-f-c1-c2} shows the results of simulations for
uncorrelated messages, and messages with 2 and 3 autocorrelation constraints\cite{simul-y-value}.
The performance of all 3 cases is similar, both in the number of iterations,
and in the distance from their respective limits\cite{performance,gzip}. 

Although we demonstrated the ability to increase the possible noise
rate the decoder can handle while keeping the rate fixed, the converse
is also possible. Keeping the channel noise fixed, higher rates are
achievable. Turning to figure \ref{fig-dig-chnl}, the rate is defined
as $R\equiv \frac{K}{N}$, but our encoder compresses the messages
as well, thus the achievable rate is actually greater. For $P_{b}=0$,
the total rate is $R^{\star }\equiv \frac{L}{N}=\frac{R}{H\left(X\right)}$
(where $H\left(X\right)$ is the entropy of the message); for $H\left(X\right)<R$
rates greater than 1 are feasible. 

The complexity of the encoding process remains linear ($O\left(N\right)$),
since the calculation of a small number of autocorrelation coefficients
is linear. The decoder's complexity per iteration scales as $\frac{N}{t}q^{2}u$
for a LDPCC decoder over $GF\left(q\right)$, with $\frac{N}{t}$
blocks, and $u$ checks per block\cite{LDPC-GF(q)}. Our decoder adds
an order of $\frac{N}{t}\left(q^{2}+q\right)$ operations for the
calculation of the block prior probabilities, calculating $q$ probabilities,
based on the pairs of $q$ states of the 2 neighboring blocks, in
equation \ref{tm-vertical-pass}. The algorithm lends itself to parallel
implementation which can reduce the complexity to $O\left(1\right)$. 

All simulations shown in this paper were done assuming BSC, however
the formulation is channel independent, and can be applied to any
channel (e.g., the popular Gaussian channel). The method can be also
applied to different codes as well (e.g., Turbo-Code), by taking into
account the dynamically updated block probabilities. 

The method described in this paper can easily be extended to higher
order correlation functions (e.g., 3-point correlation function, i.e.,
$C_{kk\prime }=\frac{1}{L}\sum _{i=1}^{L}X_{i}X_{i+k}X_{i+k\prime }$,
with periodic boundary conditions). Adding more correlations has the
effect of reducing the entropy by lifting the degeneracy associated
with 2-point correlation functions. The size of the transfer-matrix,
however, does not have to increase, thus the decoding complexity is
unchanged. For example, a transfer matrix of two bits can accommodate
1 additional autocorrelation - $C_{12}$, and a transfer matrix of
3 bits can accommodate 3 additional autocorrelations - $C_{12}$,
$C_{13}$, $C_{123}$, the last one being a 4-point correlation function\cite{symbols}. 

We have demonstrated the ability to nearly saturate Shannon's limit
without compression mechanisms. The algorithm can be used for utilizing
specific noise statistics as well. This work paves the way for additional
extensive theoretical research and various practical implementations. 

Fruitful discussions with E. Kanter are acknowledged. We would like
to thank Shlomo Shamai for critical comments on the manuscript.


\begin{thebibliography}{10}
\bibitem{prl-referee}Corrections to the leading order are of the type $\lambda _{max}^{\frac{L}{2}}\left(1+\sum _{i}\left(\frac{\lambda _{i}}{\lambda _{max}}\right)^{\frac{L}{2}}\right)$,
see for instance \cite{baxter-TM}
\bibitem{C1-gauge}The entropy of a single correlation constraint is obtained in a straightforward
fashion by performing a gauge. For $C_{1}$ the gauge $\sigma _{i}=x_{i}x_{i+1}$
gives $L$ new variables with a bias $P=\frac{1+C_{1}}{2}$, and the
entropy is given by Shannon's result, eq. \ref{Shannon-Capacity-eq}
\bibitem{corr-cluster}Note that imposing short range autocorrelation constraints, $C_{k}$,
induces long repetitions, symbols, in the sequence. For example, for
positive $C_{1}$ and $C_{2}$, one finds subsequences of the same
binary digit, whose length is much greater than 2 (the autocorrelation
range).
\bibitem{frustrated_interactions}The sign of each interaction can be positive or negative, leading
to frustration. As the entropy approaches the transition, the absolute
values of the interactions ($\left|J_{i}\right|$) diverge.
\bibitem{MN-Gallager}The range of $m$ and $n$ is specified for the MN class of LDPCC,
for Gallager-type codes $m\in \left\{ 1\ldots \frac{N-k}{t}\right\} $,
$n\in \left\{ 1\ldots \frac{N}{t}\right\} $
\bibitem{simul-y-value}The interaction strengths for the simulations shown in figure \ref{fig-pb-f-c1-c2}
are: ($y_{1}^{\star }\simeq 0.30$, $y_{2}^{\star }\simeq 0.61$)/($y_{1}^{\star }\simeq -0.31$,
$y_{2}^{\star }\simeq 0.23$, $y_{3}^{\star }\simeq -0.23$) for the
2/3 autocorrelation constraints respectively.
\bibitem{performance}The number of decoder iterations ranges from $<10$ iterations when
the noise level is low, to a few tens of iterations when the threshold
of the code is approached. The maximal noise level tolerated by the
decoder is within 10\% of $f_{\infty }$ (the bound for the noise
level for infinite length messages) in all 3 cases.
\bibitem{gzip}Given the channel characterization, $f$, the usual method would consist
of applying standard compression algorithms (e.g. compress, gzip,
bzip2), and then sending the compressed message at the appropriate
rate. For the 2 classes shown in figure \ref{fig-pb-f-c1-c2}, the
measured compression in simulations is about 80\% of the entropy (for
sequences up to $10^{6}$). Our results are superior in that they
utilize the message entropy to a greater extent, as can be seen from
figure \ref{fig-pb-f-c1-c2}.
\bibitem{symbols}The number of autocorrelation functions grows exponentially with the
block size, as do the number of symbols for a given block, which are
used in most compression techniques.
\end{thebibliography}
\end{document}